\documentclass[twocolumn,nopacs,superscriptaddress,preprintnumbers,nofootinbib,prd]{revtex4}
\usepackage{longtable}
\usepackage{amsmath,amssymb}
\usepackage{graphicx}
\usepackage{dcolumn}
\usepackage{epsfig,color}
\usepackage{bm}
\usepackage{verbatim}
\def\be{\begin{equation}}
\def\ee{\end{equation}}

\begin{document}

\title{Threshold effects in heavy quarkonium spectroscopy\footnote{Proceedings of the Symposium ``Symmetries and Order: Algebraic Methods in Many Body Systems" in honor of Professor Francesco Iachello on the occasion of his retirement, October 5-6 2018, Yale University.}}

\author{J. Ferretti}\email[]{jacopo.ferretti@yale.edu}
\affiliation{Center for Theoretical Physics, Sloane Physics Laboratory, Yale University, New Haven, Connecticut 06520-8120, USA}

\begin{abstract}
In this contribution, we discuss the possible importance of continuum-coupling (or threshold) effects in heavy quarkonium spectroscopy. 
Our calculations are carried out in a coupled-channel model, where meson-meson higher Fock (or molecular-type) components are introduced in $Q \bar Q$ bare meson wave functions by means of a pair-creation mechanism.
After providing a quick resume of the main characteristics of the coupled-channel model, we briefly discuss its application to the calculation of the masses of heavy quarkonium-like $\chi_{\rm c}(2P)$ and $\chi_{\rm b}(3P)$ states with threshold corrections.
We show that the introduction of pair-creation effects in the Quark Model (QM) formalism makes it possible to explain the deviations of $\chi_{\rm c}(2P)$ states' masses from the experimental data, without affecting the good QM description of the properties of $\chi_{\rm b}(3P)$ states.
\end{abstract}

\maketitle

\section{Introduction}
The Quark Model (QM) formalism provides a good overall description of meson and baryon observables, including the spectrum (especially the lower-energy part) \citep{Eichten:1974af,Godfrey:1985xj,Capstick:1986bm,Iachello:1991re,Bijker:1994yr,Ferraris:1995ui,Ferretti:2011zz}, the open-flavor strong decay amplitudes \citep{Bijker:1994yr,Micu:1968mk,LeYaouanc:1972vsx,Kokoski:1985is,Capstick:1993kb,Barnes:2005pb,Strong2015,Ferretti:2015rsa}, the nucleon electromagnetic form factors \cite{Iachello:1972nu,Santopinto:1998ma,Sanctis:2000eg,DeSanctis:2011zz}, and so on.
However, some difficulties emerge when one moves to higher energies, specifically to higher-lying meson (baryon) radial excitations.
One of the main problems, both from the theoretical and experimental point of view, is related to the emergence of {\it exotic} candidates. These are mesons (baryons) characterized by an unconventional (non-$q \bar q$/non-$qqq$) quark structure and/or non-standard quantum numbers. Some examples include the $X(3872)$ [now $\chi_{\rm c1}(3872)$] \cite{Choi:2003ue,Acosta:2003zx,Abazov:2004kp,Tanabashi:2018oca},  $X(4140)$ \cite{Aaltonen} and $X(4260)$ mesons \cite{Choi:2007wga}.
From a theoretical/phenomenological point of view, the previous exotic meson candidates can be described variously. Some of the main interpretations include compact tetraquark states \cite{Jaffe:1976ih,Barbour:1979qi,Weinstein:1983gd,SilvestreBrac:1993ss,Brink:1998as,Maiani:2004vq,Barnea:2006sd,Santopinto:2006my,Ebert:2008wm,Deng:2014gqa,Zhao:2014qva,Anwar:2017toa}, meson-meson molecules \cite{Weinstein:1990gu,Manohar:1992nd,Tornqvist:1993ng,Martins:1994hd,Swanson:2003tb,Hanhart:2007yq,Thomas:2008ja,Baru:2011rs,Valderrama:2012jv,Aceti:2012cb,Guo:2013sya}, the result of kinematic or threshold effects caused by virtual particles \cite{Heikkila:1983wd,Pennington:2007xr,Li:2009ad,Danilkin:2010cc,Ortega:2012rs,Ferretti:2013faa,Ferretti:2013vua,Achasov:2015oia,Kang:2016jxw,Lu:2016mbb,Ferretti:2018tco}, or hadro-quarkonia (hadro-charmonia) \cite{Dubynskiy:2008mq,Guo:2008zg,Wang:2009hi,Voloshin:2013dpa,Li:2013ssa,Wang:2013kra,Brambilla:2015rqa,Alberti:2016dru,Panteleeva:2018ijz,Ferretti:2018kzy}.
More precise experimental informations on those states, as well as a deeper understanding of the quarkonium spectrum and its patterns, will make it possible to rule out those interpretations which are not compatible with the experimental data \cite{Tanabashi:2018oca}. For recent reviews on exotics, see Refs. \cite{Esposito:2016noz,Olsen:2017bmm,Guo:2017jvc}.

In this contribution, we focus on the interpretation of the previously mentioned $X$-type exotic mesons as the result of threshold effects caused by virtual particles.
To do that, we make use of the Unquenched Quark Model (UQM) formalism with some modifications \cite{Ferretti:2018kzy}.
The UQM is an extension of the QM. It makes it possible to include the effects of virtual $q \bar q$ pairs in the na\"ive QM formalism by means of a $q \bar q$ pair-creation mechanism \cite{Micu:1968mk,LeYaouanc:1972vsx,Heikkila:1983wd,Ferretti:2012zz,Geiger:1991qe,Bijker:2009up,Bijker:2012zza}.
Above meson-meson thresholds, the creation of $q \bar q$ pairs from the vacuum is responsible of the open-flavor strong decays of the meson of interest \cite{Micu:1968mk,LeYaouanc:1972vsx,Kokoski:1985is,Barnes:2005pb,Ferretti:2015rsa}. Below threshold, it is responsible of the coupling between the meson of interest and meson-meson continuum (or molecular-type) components \cite{Heikkila:1983wd,Ferretti:2012zz,Geiger:1991qe}.
After discussing our modifications to the ``standard" version of the UQM, including some extra hypotheses to make the UQM calculations converge and remove unphysical results, we show that the introduction of meson-meson continuum components in the bare wave function of the meson of interest can provide a non-negligible correction to the meson energy and the emergence of significant continuum components in its wave function \cite{Ferretti:2018kzy}. Other possible applications of the previous coupled-channel formalism are also briefly summarized.

\section{Unquenching the quark model}
The procedure for ``unquenching the Quark Model" consists in the introduction of higher Fock components in quark-antiquark bare meson wave functions,
\begin{equation}
	\label{eqn:QQ-WF}
	\left| Q \bar Q \right\rangle \rightarrow \left| Q \bar Q \right\rangle + \left| Q \bar q - q \bar Q \right\rangle + \left| Q \bar Q g \right\rangle 
	+ ...  \mbox{ }.
\end{equation}
Here, $g$ is a constituent gluon, $\left| Q \bar q - q \bar Q \right\rangle$ a tetraquark or meson-meson molecular-type component, and $\left| Q \bar Q g \right\rangle$ a hybrid one.
The first step, namely the introduction of $\left| Q \bar q - q \bar Q \right\rangle$ components in heavy quarkonium-like meson spectroscopy, has already been carried out and some observables have been calculated \cite{Heikkila:1983wd,Pennington:2007xr,Ortega:2012rs,Ferretti:2013faa,Ferretti:2013vua,Lu:2016mbb,Ferretti:2018tco,Ferretti:2012zz,Ferretti:2014xqa}. For the introduction of $\left| Q \bar Q g \right\rangle$ components in quarkonium spectroscopy, see Ref. \cite{LeYaouanc:1984gh}.

If we restrict the extra terms of Eq. (\ref{eqn:QQ-WF}) to molecular-type components, in the Unquenched Quark Model (UQM) formalism the quarkonium-like meson wave function can be written as
\begin{equation}	
	\label{eqn:Psi-A}
	\footnotesize
	\begin{array}{l}	
	\left| \psi_A \right\rangle = {\cal N} \left[ \left| A \right\rangle + \displaystyle \sum_{BC \ell J} \int q^2 dq
	\left| BC q \ell J \right\rangle \frac{ \left\langle BC q \ell J \right| T^{\dagger} \left| A \right\rangle}{M_A - E_B - E_C} \right] ~. 
	\end{array}
\end{equation}
Here, ${\cal N}$ is a normalization factor, $\left| \psi_A \right\rangle$ is the superposition of a zeroth order quark-antiquark configuration, $\left| A \right\rangle$, plus a sum over all the possible higher Fock components, $\left| BC \right\rangle$, due to the creation of quark-antiquark pairs with vacuum quantum numbers. 
The sum is extended over a complete set of intermediate meson-meson states, $\left| BC \right\rangle$, with energies $E_{B,C} = \sqrt{M_{B,C}^2 + q^2}$; $M_A$ is the physical mass of the meson $A$; $q$ and $\ell$ are the relative radial momentum and orbital angular momentum of $B$ and $C$, and $J$ is the total angular momentum, with ${\bf J} = {\bf J}_B + {\bf J}_C + {\bm \ell}$.
The symbol $T^{\dagger}$ in Eq. (\ref{eqn:Psi-A}) stands for the pair-creation operator of Refs. \cite{Ferretti:2015rsa,Ferretti:2013faa,Ferretti:2013vua,Ferretti:2018tco,Ferretti:2012zz,Ferretti:2014xqa,Strong2015,Santopinto:2016fgs,Garcia-Tecocoatzi:2016rcj}. See also Refs. \cite{Geiger:1991qe,Bijker:2009up,Bijker:2012zza}. Below threshold, $T^{\dagger}$ is responsible of the coupling between a bare meson, $\left| A \right\rangle$, and meson-meson continuum components, $\left| BC \right\rangle$; above threshold, it is responsible of $A \rightarrow BC$ open-flavor strong decays, which proceed via the creation of a light $q \bar q$ pair (with $q = u$, $d$ or $s$) from the vacuum.
Given this, in the UQM the expectation value of a meson observable, $\hat {\mathcal O}_{\rm m}$, on the quarkonium-like states of Eq. (\ref{eqn:Psi-A}) is computed as
\begin{equation}
	\left\langle \psi_A \right| \hat {\mathcal O}_{\rm m} \left| \psi_A \right\rangle = \left\langle \hat {\mathcal O}_{\rm m} \right\rangle_{\rm val} 
	+ \left\langle \hat {\mathcal O}_{\rm m} \right\rangle_{\rm cont}  \mbox{ },
\end{equation}
where $\left\langle \hat {\mathcal O}_{\rm m} \right\rangle_{\rm val}$ and $\left\langle \hat {\mathcal O}_{\rm m} \right\rangle_{\rm cont}$ stand for the expectation values of $\hat {\mathcal O}_{\rm m}$ on the valence, $\left| A \right\rangle$, and continuum components, $\left| BC \right\rangle$, respectively.

As an example, we describe the procedure to calculate the masses of quarkonium-like states with self-energy corrections.
The physical meson masses are related to the bare and self-energies via
\begin{equation}
	\label{eqn:Ma-UQM}
	M_A = E_A + \Sigma(M_A)  \mbox{ }.
\end{equation}
Here, $E_A$ are the bare energies of the meson $A$, which have to be computed in a specific quark model; for example, we make use of the relativized QM of Ref. \cite{Godfrey:1985xj}. These energies are calculated by considering mesons as the bound states of a constituent quark-antiquark pair bounded by one-gluon-exchange forces.
$\Sigma(M_A)$ are the self-energy corrections to the bare meson masses, resulting from the coupling between the bare, $\left| A \right\rangle$, and the continuum components, $\left| BC \right\rangle$. They can be written as
\begin{equation}
	\label{eqn:self-a}
	\Sigma(M_A) = \sum_{BC} \int_0^{\infty} q^2 dq \mbox{ } 
	\frac{\left| \left\langle BC q \ell J \right| T^\dag \left| A \right\rangle \right|^2}{M_A - E_B - E_C}  \mbox{ },
\end{equation}
where the sum is extended over a complete set of intermediate meson-meson states $BC$.

One can also calculate the norm of the continuum (or molecular-type) component of a quarkonium-like state via \cite{Ferretti:2018tco,Ferretti:2012zz}
\begin{equation}
	\label{eqn:Pa-sea}
	P_A^{\rm sea} = \sum_{BC\ell J} \int_0^\infty q^2 dq \mbox{ } 
	\frac{\left| \left\langle BC q  \, \ell J \right| T^\dag \left| A \right\rangle \right|^2}{(M_A - E_B - E_C)^2}  \mbox{ },
\end{equation}
where the probability to find the meson in its valence component, $P_A^{\rm val}$, is given by $P_A^{\rm val} = 1 - P_A^{\rm sea}$.

The UQM formalism has been extensively used in the past to compute both baryon and meson observables, including the calculation of baryon \cite{Garcia-Tecocoatzi:2016rcj,SilvestreBrac:1991pw,Morel:2002vk} and meson \cite{Heikkila:1983wd,Pennington:2007xr,Ferretti:2013faa,Ferretti:2013vua,Lu:2016mbb,Ferretti:2018tco,Ferretti:2012zz} masses with self-energy corrections, heavy quarkonium hidden flavor strong decays \cite{Ferretti:2018tco}, and the strangeness contribution to the nucleon electromagnetic form factors \cite{Bijker:2012zza,Geiger:1996re}.
Despite of its merits, including its simplicity and versatility, the UQM calculations do not converge quickly. 
Indeed, it can be easily shown that, as the tower of meson-meson intermediate states $\left| BC \right\rangle$ is enlarged, the contribution of continuum or sea components to hadron observables keeps growing larger and larger.
Below, we discuss a simple procedure to ``renormalize" the UQM results. More details can be found in Ref. \cite{Ferretti:2018tco}.

\section{A COUPLED-CHANNEL MODEL FOR HEAVY QUARKONIUM-LIKE STATES}
After discussing the main features of the UQM formalism for mesons \cite{Ferretti:2015rsa,Ferretti:2013faa,Ferretti:2013vua,Ferretti:2012zz,Ferretti:2014xqa}, here we show a procedure to ``renormalize" it and avoid the production of unphysical results \cite{Ferretti:2018tco}.
In particular, as a first step, we give a brief resume of a simple coupled-channel method to compute the physical masses of quarkonium-like mesons, $M_A$, with threshold corrections \cite{Ferretti:2018tco}.
The method is based on the UQM formalism, plus the following hypotheses and prescriptions: 
a) The method is not used to perform a global fit to the heavy quarkonium spectrum, but it is applied only to specific meson multiplets, like $\chi_{\rm c}(2P)$ and $\chi_{\rm b}(3P)$; b) Only the closest complete set of accessible SU(N)$_{\rm flavor} \otimes$ SU(2)$_{\rm spin}$ open-flavor meson-meson intermediate states (e.g. $1S1S$, $1S1P$ or $1S2S$) can influence the multiplet structure.
The other (lower or upper) meson-meson thresholds, which are further in energy, are supposed to give some kind of global or background contribution, which can be subtracted; c) The presence of a certain complete set of open-flavor intermediate states does not affect the properties of a single resonance, but it influences those of all the multiplet members.
Thus, the net effect of the intermediate states on a quarkonium-like meson multiplet is similar to that of a spin-orbit or hyperfine splitting.

\begin{table*}
\caption{Comparison between the experimental masses \cite{Tanabashi:2018oca} of $\chi_{\rm c}(2P)$ and $\chi_{\rm b}(3P)$ states and theoretical predictions from Ref. \cite{Ferretti:2018tco}. The bare masses, $E_A$, are extracted from the original relativized QM fit of Refs. \cite{Godfrey:1985xj,Barnes:2005pb,Godfrey:2015dia}. The experimental results denoted by $\dag$ are extracted from Ref. \cite{Godfrey:2015dia}, where the authors used predicted multiplet mass splittings in combination with the measured $\chi_{\rm b1}(3P)$ mass. In the $h_{\rm c}(2P)$ case, we use the same value for the physical mass as the bare one \cite{Godfrey:1985xj}.}
\label{tab:ChiC(2P)-splittings}
\tabcolsep7pt\begin{tabular}{lcccc}
\hline
State  & $E_A$ [MeV]  & $\Sigma(M_A) - \Delta$ [MeV] & $M_A^{\rm th}$ [MeV] & $M_A^{\rm exp}$ [MeV] \\
\hline
$h_{\rm c}(2P)$       & 3956               & $-16$                                      & 3940                            & -- \\
$\chi_{\rm c0}(2P)$ & 3916               & 0                                             & 3916                            & 3918  \\
$\chi_{\rm c1}(2P)$ & 3953               & $-65$                                      & 3888                            & 3872  \\
$\chi_{\rm c2}(2P)$ & 3979               & $-30$                                      & 3949                            & 3927  \\ 
&&&&  \\                              
$h_{\rm b}(3P)$       & 10541            & $-4$                                         & 10538                          & 10519$^\dag$ \\
$\chi_{\rm b0}(3P)$ & 10522             & 0                                              & 10522                          & 10500$^\dag$  \\
$\chi_{\rm b1}(3P)$ & 10538             & $-2$                                         & 10537                          & 10512  \\
$\chi_{\rm b2}(3P)$ & 10550             & $-7$                                         & 10543                          & 10528$^\dag$  \\ 
\hline   
\end{tabular}
\end{table*}

\subsection{Threshold mass-shifts}
Under the previous hypotheses, the physical masses of the members of a quarkonium-like meson multiplet are computed as \cite{Ferretti:2018tco}
\begin{equation}
	\label{eqn:new-Ma}
	M_A = E_A + \Sigma(M_A) + \Delta \mbox{ },
\end{equation}
where $E_A$ is the bare mass of meson $A$, whose value is extracted from the relativized QM predictions of Refs. \cite{Godfrey:1985xj,Barnes:2005pb,Godfrey:2015dia}.
It is worth noting that here and in Ref. \cite{Ferretti:2018tco}, contrary to the calculations of Refs. \cite{Ferretti:2013faa,Ferretti:2013vua}, the relativized QM parameters are not fitted to the reproduction of the physical masses of Eq. (\ref{eqn:Ma-UQM}). The bare mass values are directly extracted from the original relativized QM fit of Ref. \cite{Godfrey:1985xj}. See also Refs. \cite{Barnes:2005pb,Godfrey:2015dia}.
The second term in Eq. (\ref{eqn:new-Ma}) is the self-energy correction of Eq. (\ref{eqn:self-a}). 
In the case of heavy quarkonium-like states around the opening of the first meson-meson decay thresholds, the closest complete set of meson-meson intermediate states is made up of $1S1S$ open-flavor mesons. 
For example, in the case of the $\chi_{\rm c}(2P)$ multiplet, we consider $D \bar D$, $D \bar D^*$, $D^* \bar D^*$, $D_{\rm s} \bar D_{\rm s}$, $D_{\rm s} \bar D_{\rm s}^*$, $D_{\rm s}^* \bar D_{\rm s}^*$, $\eta_{\rm c} \eta_{\rm c}$, $\eta_{\rm c} J/\psi$, and $J/\psi J/\psi$ meson-meson components \cite{Ferretti:2018tco}. 

The pair-creation model parameters, which we need in the calculation of the $\left\langle BC q \ell J \right| T^\dag \left| A \right\rangle$ vertices of Eq. (\ref{eqn:self-a}), were fitted to the open-flavor strong decays of charmonia \cite[Table II]{Ferretti:2013faa} and bottomonia \cite[Table I]{Ferretti:2013vua}; see also \cite[Table II]{Ferretti:2015rsa}.
Therefore, for each multiplet, there is only one free parameter, $\Delta$.
This is the smallest self-energy correction (in terms of absolute value) to the bare mass of a multiplet member; see \cite[Sec. 2]{Ferretti:2018tco} and the following section.

\section{THRESHOLD MASS-SHIFTS IN $\chi_{\rm c}(2P)$ and $\chi_{\rm b}(3P)$ MULTIPLETS}
We calculate the threshold mass shifts of the $\chi_{\rm c}(2P)$ and $\chi_{\rm b}(3P)$ multiplet members due to a complete set of ground state $1S 1S$ meson loops, like $D \bar D$, $D \bar D^*$ ($B \bar B$, $B \bar B^*$), and so on \cite{Ferretti:2018tco}.
The values of the bare masses, $E_A$, are extracted from the relativized model \cite{Godfrey:1985xj}, those of the physical masses, $M_A$, from the PDG \cite{Tanabashi:2018oca}.
The self-energy corrections, $\Sigma(M_A)$, are computed according to Eq. (\ref{eqn:self-a}), using the same pair-creation model parameter values as Refs. \cite{Ferretti:2015rsa,Ferretti:2013faa,Ferretti:2013vua}. 

\begin{figure}[h]
\centerline{\includegraphics[width=200pt]{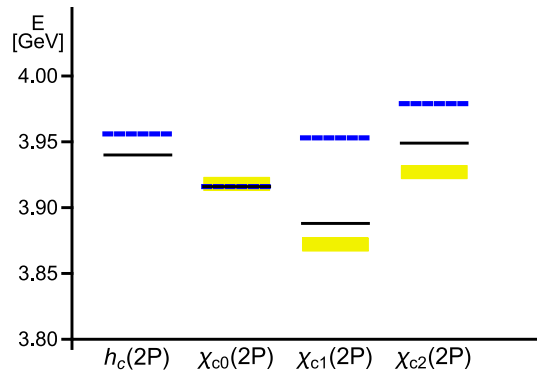}}
\caption{$\chi_{\rm c}(2P)$ multiplet: masses with threshold corrections. Yellow boxes, blue dashed and black continuous lines correspond to the experimental \cite{Tanabashi:2018oca}, calculated bare and physical masses, respectively. The calculated bare and physical mass values are extracted from Table \ref{tab:ChiC(2P)-splittings}; see also \cite[Fig. 1 and Table 1]{Ferretti:2018tco}.}
\label{fig:ChiC(2P)-splittings}
\end{figure}

In the case of the $\chi_{\rm c}(2P)$ multiplet, we get: $\Sigma(M_{h_{\rm c}(2P)}) = -119$ MeV, $\Sigma(M_{\chi_{\rm c0}(2P)}) = -103$ MeV, $\Sigma(M_{\chi_{\rm c1}(2P)}) = -168$ MeV, $\Sigma(M_{\chi_{\rm c2}(2P)}) = -133$ MeV. Thus, $\Delta = \Sigma(M_{\chi_{\rm c0}(2P)})$.
In the case of the $\chi_{\rm b}(3P)$ multiplet, we obtain: $\Sigma(M_{h_{\rm b}(3P)}) = -116$ MeV, $\Sigma(M_{\chi_{\rm b0}(3P)}) = \Delta = -112$ MeV, $\Sigma(M_{\chi_{\rm b1}(3P)}) = -114$ MeV, $\Sigma(M_{\chi_{\rm b2}(3P)}) = -119$ MeV. 
The values of the calculated physical masses of the $\chi_{\rm c}(2P)$ and $\chi_{\rm b}(3P)$ multiplet members via Eq. (\ref{eqn:new-Ma}) are reported in Table \ref{tab:ChiC(2P)-splittings}. See also Fig. \ref{fig:ChiC(2P)-splittings}.
It is worth noting that: I) Our theoretical predictions agree with the data within the typical error of a QM calculation ($\sim 30-50$ MeV); II) Among the $\chi_{\rm c}(2P)$ multiplet members, the $\chi_{\rm c1}(2P)$ receives the largest contribution from the continuum. This continuum contribution is necessary to lower the relativized QM prediction, 3.95 GeV, towards the observed value of the meson mass, 3871.69 MeV \cite{Tanabashi:2018oca}; III) In the $\chi_{\rm c}(2P)$ case, threshold effects break the usual mass pattern of a $\chi$-type multiplet, namely $M_{\chi_0} < M_{\chi_1} \approx M_{\rm h} < M_{\chi_2}$; IV) The threshold effects are negligible in the $\chi_{\rm b}(3P)$ case. Because of this, we interpret $\chi_{\rm b}(3P)$ states as (almost) pure bottomonia; V) Unlike the $\chi_{\rm c}(2P)$ case, the usual mass pattern within a $\chi$-type multiplet, namely $M_{\chi_0} < M_{\chi_1} \approx M_{\rm h} < M_{\chi_2}$, in the $\chi_{\rm b}(3P)$ case is now respected.  

\section{CONCLUSIONS}
We discussed the possible importance of continuum-coupling (or threshold) effects in heavy quarkonium spectroscopy. 
Our calculations were carried out in a coupled-channel model, where meson-meson higher Fock (or molecular-type) components were introduced in $Q \bar Q$ bare meson wave functions by means of a pair-creation mechanism.
After providing a quick resume of the main characteristics of the coupled-channel model, we briefly discussed its application to the calculation of the masses of heavy quarkonium-like $\chi_{\rm c}(2P)$ and $\chi_{\rm b}(3P)$ states with threshold corrections.
Our results are compatible with the present experimental data \cite{Tanabashi:2018oca}.

\begin{acknowledgements}
This work was supported by US Department of Energy Grant No. DE-FG-02-91ER-40608.
\end{acknowledgements}

\end{document}